\documentstyle[floats,aps]{revtex}
\begin{document}
\draft
\catcode`\@=11
\catcode`\@=12
\twocolumn[\hsize\textwidth\columnwidth\hsize\csname%
@twocolumnfalse\endcsname

\title{Clustering dynamics in globally coupled map lattices}

\author{Fagen Xie$^{1,2}$ and Gang Hu$^{1,3}$}
\address{
$^1$ China Center of Advanced Science and Technology, (World Laboratory), 
P.O. Box 8730, Beijing, 100080, China
\\$^2$ Institute of Theoretical Physics, Academia Sinica, Beijing 100080, China
\\$^3$ Department of Physics, Beijing Normal University, Beijing 100875, China}

\date{Received: \today}

\maketitle

\begin{abstract}

Clustering bifurcations are investigated by considering models of globally 
coupled map lattices. Typical classes of clustering bifurcations are revealed. 
The clustering bifurcation thresholds of the coupled system are closely 
related to the bifurcation structures of single map.  In particular, 
cluster-doubling bifurcation induced period-doubling bifurcations and 
clustering induced chaos are found. At the onset of multiple-cluster 
states, equal-site-occupation-partition, and consequently, equal-phase-shift 
states \char '133the so-called antiphase states reported previously, 
 Phys. Rev. Lett. {\bf 65}, 1749 (1990)\char '135~are always identified 
 numerically. 

\end{abstract}

\pacs{PACS numbers: 05.45.+b}]

\narrowtext
\tighten

The investigation of globally coupled extended systems has attracted a 
rapidly growing interest in recent years$^{1-6}$. They arise naturally in 
studies of Josephson junctions arrays$^1$,  multimode laser$^2$, 
charge-density wave$^3$, oscillatory neuronal systems$^4$ and so on. 
A number of intriguing  and novel high-dimensional features have been 
revealed in these spatiotemporal systems. For instance, 
a curious and interesting dynamical state, so called {\it antiphase} state has 
been revealed both numerically and experimentally$^{1,2}$.  
Such a state is periodic in time, with each element of the system 
having precisely the same wave form. However, the motion of each element is 
just shifted by certain phase from its ``neighbor". 
This state is closely related to the {\it clustering}, 
which is numerically studied in the globally coupled chaotic maps$^5$. 
The antiphase state is a {\it clustering} with equal occupation elements in each 
cluster.  Up to date, the mechanism underlying this fascinating phenomenon has 
not been revealed. In this letter we will thoroughly analyze the  
bifurcation mechanisms and phase diagrams of clusterizations and reveal 
the interesting general features of equal-site-occupation-partition (ESOP) 
 and equal-phase-shift (EPS) states at bifurcation points of clustering. (Here 
we call the antiphase state as EPS state 
because multiple phases may appear in clustering bifurcations.) We take the 
following globally coupled map lattice (GCML) 

 \begin{eqnarray}
 x_{n+1}(i)=(1-\epsilon)f[x_{n}(i)]+\displaystyle{\epsilon\over L}
 \sum_{j=1}^Lf[x_{n}(j)],~ i=1,\cdots,L,
 \label{e1} 
 \end{eqnarray}
 as our model, 
where $n$ denotes the discrete time, $i$ labels the lattice site with 
$L$ system size. $f(x)$ prescribes the local dynamics, and is chosen 
as the logistic map $f(x)=ax(1-x)$. $\epsilon$ gives the long-range 
coupling strength. In Refs. 5, Kaneko {\it et al} presented very 
rich and interesting behaviors of (\ref{e1}) for positive $\epsilon$. 
 Negative $\epsilon$ represents also many 
practical situations, such as antiferromagnetic coupling$^6$ and resistance 
coupling and so on. Therefore, it is useful to unify the investigations of 
Eq. (\ref{e1}) for both positive and negative $\epsilon$.

First, we consider clustering bifurcations from the simplest spatially 
homogeneous configuration, so called {\it coherent} state. 
 After some simple algebra, the critical stability condition of this 
coherent state can be explicitly shown as  
\begin{eqnarray}
\epsilon_c =1 - e^{-\lambda_0},
\label{e3}
\end{eqnarray}
where $\lambda_0$ is the Lyapunov exponent of the single  logistic 
map [ note, (\ref{e3}) is generally valid for any coherent state, whatever periodic
or chaotic]. This critical stability boundary is shown in Fig. 1(a) with solid line. 
 As $\epsilon > \epsilon_c$, the coherent state always exists,
 and is {\it locally} stable, while below the solid lines, the coherent state loses 
 its stability and bifurcates to multi-cluster state. From (\ref{e3}) it 
 is clear that, coherent periodic motions are always stable for positive $\epsilon$, 
 they can lose stability only in the negative coupling regions. However, it should 
 be emphasized that many attractors may coexist with the coherent state in the 
 regime above the solid line for large system size $L$ and large $a$.
 
A class of interesting states are multi-cluster states with ESOP (i.e., 
$k$ clusters $N_1=\cdots=N_k$, with $N_i, i=1,\cdots,k$ being the occupation 
numbers of $i$th cluster), and in case of periodic motion,  each cluster 
may have the same motion except some EPS. It will be shown that 
this kind of states appear naturally at bifurcation thresholds. Afterwards, a 
period-$m$ state with $k$ clusters will be called TmCk state. It often happens 
that $m=k$, the evolving dynamics of the TkCk state can be much reduced as
\begin{eqnarray}
x_{n+1}=(1-\epsilon)f(x_n)+\displaystyle{\epsilon\over k} \sum_{j=1}^kf(x_j).
\label{e4}
\end{eqnarray}
For $k=2$, the solutions of Eq. (\ref{e4}) read 
\begin{eqnarray}
\hspace*{-0.3cm} x_{1,2}=\displaystyle\frac{1+a-a\epsilon\pm\sqrt{(1+a-a\epsilon)^2-
2(2-\epsilon)(1+a-a\epsilon)}}{2a(1-\epsilon)}.
\label{e5}
\end{eqnarray}

The stability conditions of the ESOP TkCk state can be analytically given by 
computing the products of Jacobi matrices of system (\ref{e1}).  
 After certain matrix operations, the stability analysis of system (\ref{e1}) 
 can be much simplified to a block-diagonal form of the linear matrix  
 of (\ref{e4}) as
\begin{equation}
J=\left (
\begin{array}{cc}
\displaystyle\prod_{n=0}^kM_n & 0\\
 0  & M'
\end{array}
\right)
\end{equation}
 where 
\[ 
\hspace*{-0.6cm}M_n=\left (
\begin{array}{cccc}
(1-\frac{(k-1)\epsilon}{k})f_n^1 & \frac{\epsilon}{k}f_n^2 & 
\cdots & \frac{\epsilon}{k}f_n^k  \\
\frac{\epsilon}{k}f_n^1 & (1-\frac{(k-1)\epsilon}{k})f_n^2 &
\cdots & \frac{\epsilon}{k}f_n^k  \\
\cdots &\cdots &\cdots \\
\frac{\epsilon}{k}f_n^1 & \frac{\epsilon}{k}f_n^2 &
\cdots & (1-\frac{(k-1)\epsilon}{k})f_n^k \\ 
\end{array}
\right)
\]
with $f_n^i=ax_{n+i}(1-x_{n+i})$, $i=1,2,\cdots,k$, and $x_{n+k}=x_n$. 
$M'=(1-\epsilon)^ka^k\displaystyle\prod_{n=0}^kx_n(1-x_n){\bf I}$, 
${\bf I}$ is the $(L-k)\times(L-k)$ unit matrix. Therefore, we obtain 
$L-k$ degenerate eigenvalues 
$\lambda=(1-\epsilon)^ka^k\displaystyle\prod_{n=0}^kx_n(1-x_n)$,
 and other $k$ eigenvalues. If the absolute values of all eigenvalues of $J$ 
 are less than {\it one}, the reference $k$-cluster state is stable. For $k=2$, 
The stability boundaries can be given explicitly. Increasing $a$ from small 
value, the T2C2 state appears from the spatially homogeneous state via 
saddle-node bifurcation (SN) and pitch-fork bifurcation (PK) for $\epsilon>0$ 
and $\epsilon<0$, respectively, at the following critical thresholds
\begin{equation}
\begin{array} {lrll}
{\rm\bf SN:} & a_c=&\displaystyle 1+\sqrt{1+\frac{3}{(1-\epsilon)^2}}, &\epsilon>0,\cr
{\rm\bf PK:} & a_c=& 2+\displaystyle\frac{1}{1-\epsilon}, &\epsilon<0,
\end{array}
\end{equation}
and loses its stability via Hopf bifurcation (HF) and pitch-fork bifurcation at  
\begin{equation}
\begin{array} {lrll}
{\rm\bf HF:} & a_c=& 1+\displaystyle\sqrt{1+\frac{5-3\epsilon}{(1-\epsilon)^2}}, 
& \epsilon>0,\cr
{\rm\bf PK:} & a_c=& 1+\displaystyle\frac{\sqrt{2+(\epsilon-2)^2}}{1-\epsilon},  
& \epsilon<0,
\end{array}
\end{equation}

All these bifurcation lines and other clustering bifurcation lines in the 
period-doubling region are shown in Fig. 1(b). It is remarkable that in 
Fig. 1(b) one can find a clear rule to describe the entire clustering 
bifurcation structure in $\epsilon-a$ plane from the bifurcation points in the 
$a$ axis at $\epsilon=0$. Actually, all bifurcation points for a single cell 
(at $\epsilon=0$) are multicodimension bifurcation points in the $\epsilon-a$ 
parameter plane. For $\epsilon>0$ one can find first order saddle-node 
bifurcation and second order Hopf bifurcation curves, while for $\epsilon<0$ one 
can find second order pitch-fork bifurcation and first order subcritical bifurcation 
(SC) curves, all these bifurcation curves intersect with the $a$ axis ($\epsilon=0$) 
at the critical points for single cell. These beautiful bifurcation trees can be also 
found in chaos region associated with each periodic window. 
 The enlarged regions of the rectangles in Fig. 1(a) for the bifurcations to 
  ESOP three-cluster, and five-cluster states are shown in 
  Fig.1 (c) and (d), respectively. The bifurcation figures are similar to Fig. 1(b). 
  A difference  of these figures from the two-cluster state is that  
  for $\epsilon<0$ these multi-cluster states appear from chaotic motions 
  via  saddle-node bifurcation rather than pitch-fork bifurcation. 

To give clear pictures about the clustering bifurcations we show the asymptotic 
states of system (\ref{e1}) in Figs. 2(a) and (b), where all stable homogeneous states 
are plotted by diamonds, stable multicluster states by solid lines,  
unstable states (both homogeneous and inhomogeneous) by dashed lines. 
The black regions represent stable quasiperiod motion. 
 In Fig. 2(a) we fix $\epsilon=0.2$, the system has only the 
 coherent state below $a_c\approx 3.386$ (T1C1 for $a<3$ and T2C1 for $a>3$), 
 the ESOP T2C2 state occurs via a saddle-node bifurcation at $a_c$, 
 then the two states (T2C1 and T2C2) coexist in certain $a$ interval. 
 As $a$ continuously increases, the coherent state undergoes a series of 
 period doubling bifurcations leading to chaos, and then loses coherence
  at $a\approx 3.640$. The T2C2 state subjects to Hopf bifurcation at 
 $a\approx 3.808$. In Fig. 2(b) we fix $\epsilon=-0.15$, the bifurcations 
 are essentially different from those of (a).  The T1C1 state first undergoes a 
 cluster-period-doubling bifurcation at $a\approx 2.870$ to create a stable 
 T2C2 state. After $a>3.07$, the coherent T2C1 state turns to be stable via 
 subcritical bifurcation. In a large interval $3.4>a>3.07$ the coherent T2C1 
 state coexists with multicluster state, while the T2C2 state undergoes further 
 cluster-period-doubling bifurcation  and Hopf bifurcation leading to chaos. 
 At $a\approx 3.4$ the T2C1 state bifurcates via cluster-period-doubling to 
 form a T4C2 state. Fig. 2(b) is extremely interesting due to the following novel 
 features. First, we find a cluster-doubling induced period-doubling. 
 The value $a\approx 2.870$ is far below the period-doubling condition for a 
 single map. Global coupling leads to cluster doubling at this parameter, that 
 induces period doubling in time. Second, we find a cluster-doubling sequence 1-2-4
 (and the induced period-doubling sequence). We expect that this cluster doubling 
  cascade may proceed to a very large numbers of cluster. In our case this
 cascade is stopped at $k=4$ by Hopf bifurcation at $a\approx 3.316$. 
 Nevertheless, the tendency of cluster increasing bifurcations leading to chaos can 
 be still seen in Fig. 2(c) for $a<3.4$, where we plot the number of clusters vs. $a$ for 
 the states of Fig. 2(b). It is found in Fig. 2(b) that chaos can appear for $a<3.4$, 
 where the nonlinear parameter $a$ is far below the value for chaotic motion 
 for the single map. In the same time in Fig. 2(c) we find the number of clusters 
 diverges (to the order of $L$) in this chaos region. Then we conclude this 
 chaos is made possible by clusterization, and such is clustering induced chaos. 
  It is remarkable that all analytical predictions in Fig. 1(b) are perfectly 
  confirmed by numerical simulations of (\ref{e1}) in Figs. 2.
  
 In the above we focused on the discussion of ESOP and EPS multiple cluster states. 
 On one hand, these states can be easily treated by analyzing Eqs. (\ref{e4}). 
 On the other hand, these kinds of states appear generically and naturally 
 under bifurcation conditions. For instance, in Fig. 2(a) around at $a\approx 3.386$ 
 and from arbitrary initial conditions we can get T2C1 or T2C2 
 states. Whenever we get T2C2 state it must be an ESOP state 
 ($N_1=N_2=\frac{L}{2}$ if $L$ is even, or $N_1=\frac{L-1}{2}, 
 N_2=\frac{L+1}{2}$ if $L$ is odd). In Fig. 2(b) we run Eqs. ({\ref{e1}) from 
 random initial conditions at $a=2.8$, then compute Eqs. ({\ref{e1}) by 
 gradually increasing $a$ and by using the final state for the previous 
 $a$ as the initial state for the new $a$, we can surely get  
 ESOP  and ESP states for all cluster-doubling cascade. 
 
 let us take a two-cluster state as an example. Usually, many two-cluster 
 attractors with different $N_1$ and $N_2$ ($N_1+N_2=L$) may coexist. 
 However, at the onset of two-cluster state we always first find the ESOP state. 
  In Fig. 3(a) and (b) we fix $\epsilon=0.2, L=200$, and plot the two-cluster 
  states for the occupation numbers $N_1=N_2=100$, and $N_1=98, N_2=102$, 
  respectively. It is really striking that in (b) by a slightly breaking the 
  occupation balance the threshold for two-cluster state is 
  considerably raised. In Fig. 3(c) we fix $\epsilon=-0.3, L=200$, and plot 
  $N_m$ vs. $a$, where $N_m$ is the possible smallest occupation number for the 
  two-cluster states of Eqs. (1). Small noises are added to the system sites for 
  wiping out glass states with extremely small basins. 
  It is interesting that $N_m$ goes to $\frac{L}{2}$ at both left and right 
  critical points. (This Phenomenon is also found by Kaneko {et al} for 
  $\epsilon>0$.) We find that the ESOP state naturally appears 
  at both first and second order bifurcation points.
  
  Similar behavior happens also for general $k$-cluster bifurcations. 
  In Figs. 4 and 5 we plot three and five cluster states, respectively, 
  arising via saddle-node bifurcation from spatiotemporal chaos. 
  In the two cases, the ESOP states arise much earlier than those with 
  slight deviations from equal occupation partition for both $\epsilon >0$ 
  and $\epsilon <0$. We have also examined many $k$-cluster bifurcations and 
  always find the same behavior. 
  
  In conclusion we have revealed a detailed bifurcation structure for clusterization 
  and found a cluster-doubling sequence and  clustering induced spatiotemporal 
  chaos. At any onset of $k$-cluster state one always finds ESOP and EPS state. 
  This is a general extension of the results obtained 
  by Wiesenfeld {\it et al}. However, in our case the natures of bifurcations and 
  the links of clustering bifurcations with the bifurcations of single site are 
  clearly shown  in Figs. 1 and 2, then these ESOP and EPS states of globally 
  coupled systems can be predicted, based on the bifurcation structure of single site.
  
  In this letter we took the globally coupled map lattices as our models. However, 
  the ideas can be extended to more general globally coupled systems. 
  For instance, we have examined globally coupled Josephson Junctions, the
  clustering bifurcation features are qualitatively the same as those for the 
  coupled map lattice systems. 

 \vspace*{0.2in}  
 
 This work is supported by the National Natural Science foundation of China and 
 Project of Nonlinear Science.


\begin {description}

\item{FIG. 1} Bifurcation figure for homogeneous (coherent) state. Below solid 
curves the homogeneous state loses its stability. (b) Clustering bifurcations 
in period-doubling region. SN, HP, PK and SC represent saddle-node, Hopf, 
pitch-fork and subcritical bifurcations, respectively. (c)  The blow up of the 
second rectangle of (a). The same as (b) with period-three window is considered. 
(d) The blow up of the first small rectangle of (a). The same as (c) with 
period-five window is considered.

\item{FIG. 2} Bifurcation sequences for certain couplings. Diamonds, solid lines
 and dashed lines represent stable coherent states, stable multiple cluster states
 and unstable states, respectively. (a) $\epsilon=0.2$, at $a\approx 3.386$  
 a saddle-node bifurcation for two cluster state occurs. (b) $\epsilon=-0.15$, 
 a cluster-doubling cascade induced period-doubling cascade leading to chaos is 
 shown. Note the subcritical bifurcation for T2C1 state at $a=3.070$, and the 
 Hopf bifurcation at $a=3.316$ which makes the cluster-doubling cascade incomplete.
  (c) The number of cluster, $k$, plotted vs. $a$ with $\epsilon=-0.15$. In the 
  region of clustering induced chaos $k$ is of order $L$.
 
\item{FIG. 3} T2C2 states for $L=200$. (a) $\epsilon=0.2$, $N_1=N_2=100$.
 (b) $\epsilon=0.2$, $N_1=98, N_2=102$. T2C2 state appears much earlier in 
 (a) than in (b). (c) $\epsilon=-0.3$, $L=200$. The number of the smallest site 
 population $N_m$ plotted vs. $a$. At the two bifurcation points $N_m$ goes to 
 $\frac{L}{2}$, indicating ESOP state. 
 
 \item{FIG. 4} Three-cluster states, $L=300$. (a) $\epsilon=0.05$, 
 $N_{1,2,3}=100$. (b) $\epsilon=0.05$, $N_1=98, N_2=100, N_3=102$. 
 (c) $\epsilon=-0.1$, $N_{1,2,3}=100$. (d) (b) $\epsilon=-0.1$, 
 $N_1=98, N_2=100, N_3=102$. At the onset of three-cluster state one can 
 see only the ESOP state.
 
 \item{FIG. 5} Five-cluster states, $L=500$, the same behavior as in Fig. 4. 
 (a) $\epsilon=0.005$, $N_{1,2,3,4,5}=100$. (b) $\epsilon=0.005$, $N_1=94, 
 N_{2,3,4}=100$,$N_5=106$. (c) $\epsilon=-0.005$, $N_{1,2,3,4,5}=100$. 
 (d) $\epsilon=-0.005$,  $N_1=94, N_{2,3,4}=100$,$N_5=106$.
 
\end{description}


\vspace*{-0.2in}
\begin{references}
\vspace*{-0.6in}

\bibitem{kw89} K. Wiesenfeld and P. Hadley, Phys. Rev. Lett. 
{\bf 62}, 1335 (1989); K. Y. Tsang and K. Wiesenfeld, Appl. Phys. 
Lett. {\bf 56}, 495 (1990); S. Watanabe and S. H. Strogatz, Phys. Rev. 
Lett. {\bf 70}, 2391 (1993); S. H. Strogatz and R. E. Mirollo, Phys. 
Rev. E {\bf 47}, 220 (1993); A. A. Chernikov and G. Schmidt, Phys. 
Rev. E {\bf 52}, 3415 (1995); H. G. Winful and L. Rahman, Phys. Rev. 
Lett. {\bf 65}, 1575 (1990); K. Wiesenfeld and J. W. Swift, Phys. 
Rev. E {\bf 51}, 1020 (1995); D. Dom\'{\i}nguez and H. A. Cerdeira, 
Phys. Rev. Lett. {\bf 71}, 3359 (1993); B {\bf 52}, 513 (1995). 

\bibitem{kw90} K. Wiesenfeld, C. Bracikowski, G. James, and R. Roy, 
 Phys. Rev. Lett. {\bf 65}, 1749 (1990).

\bibitem{hs91} H. Sompolinsky, D. Golomb, and D. Kleinfeld, Phys. Rev. A {\bf 43}, 6990 (1991).

\bibitem{sh89} S. H. Strogatz, C. M. Marcus, R. M. Westervett, and R. E. Mirollo, Physica D 
{\bf 36}, 23 (1989).

\bibitem{kk89} K. Kaneko, Phys. Rev. Lett. {\bf 63}, 219 (1989); 
{\bf 65}, 1391 (1989); Physica D {\bf 41}, 137 (1990); {\bf 54}, 5 (1991).

\bibitem{as95} M. Antoni and S. Ruffo, Phys. Rev. E {\bf 52}, 2361 (1995). 

\end{references}
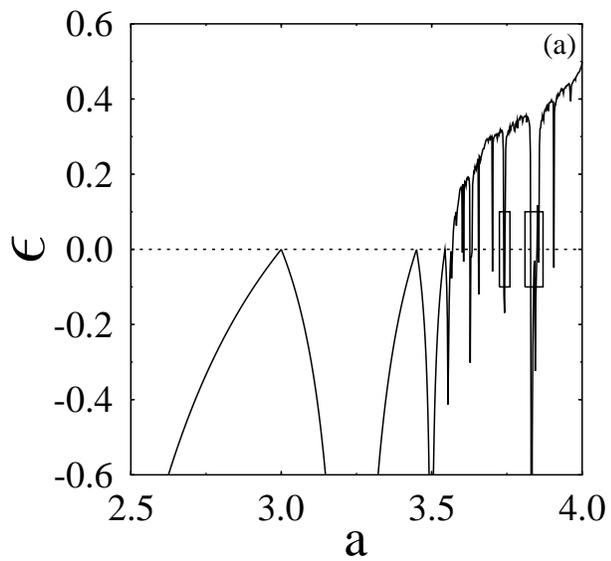
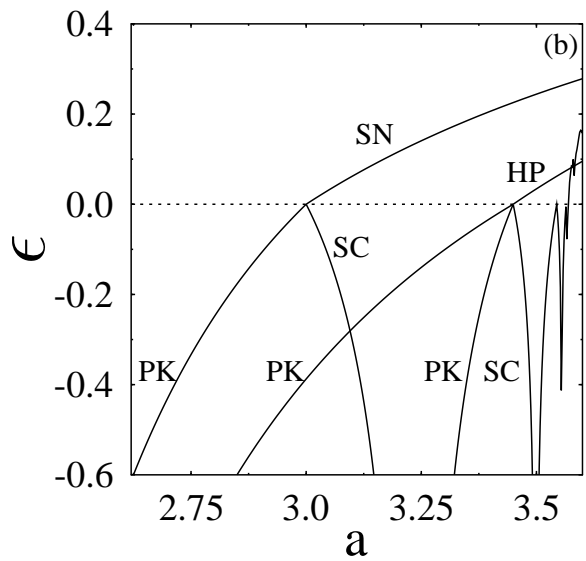
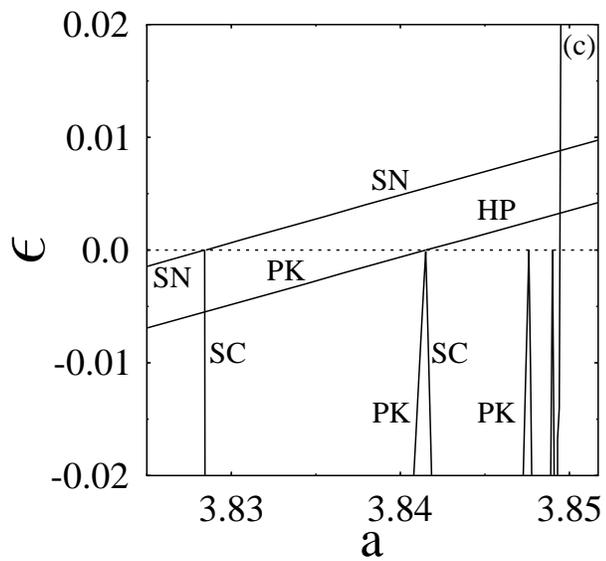
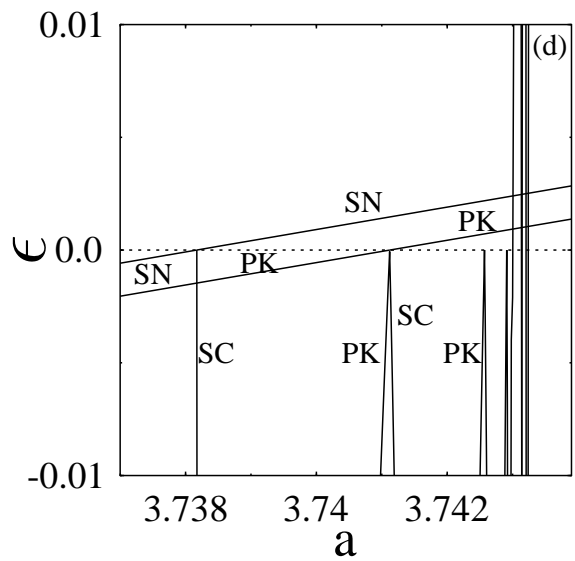
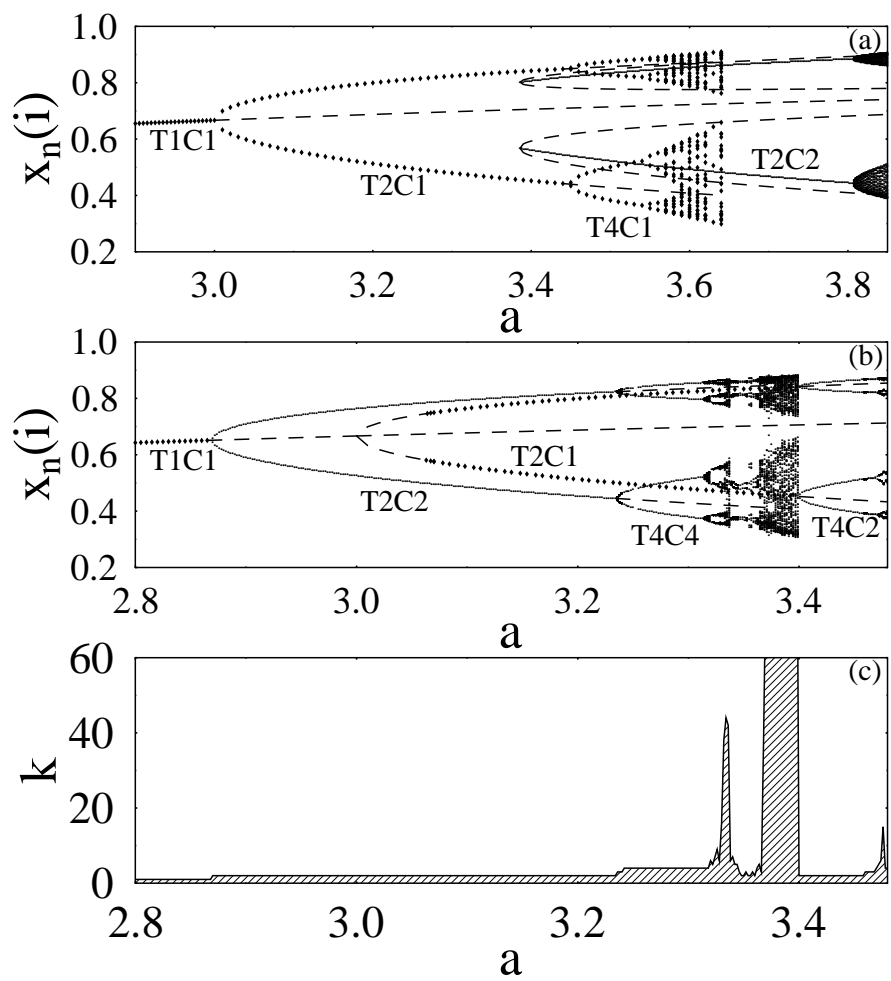
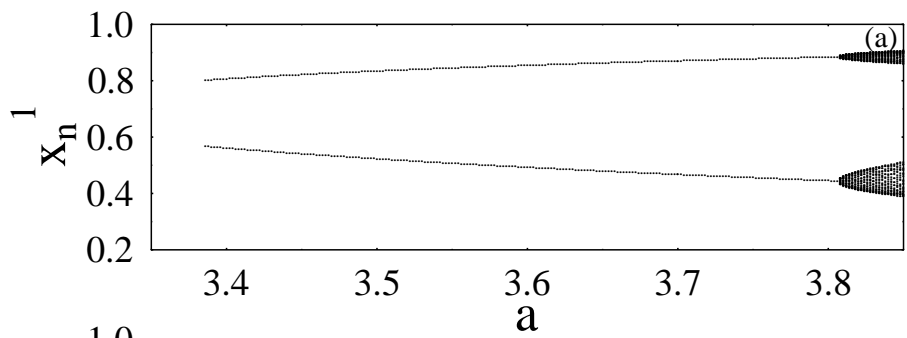
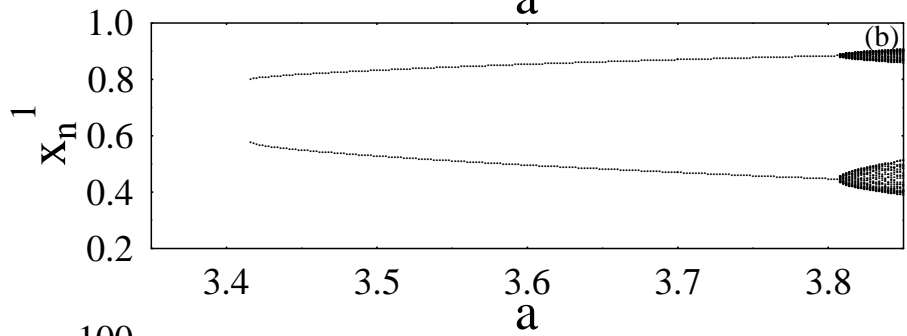
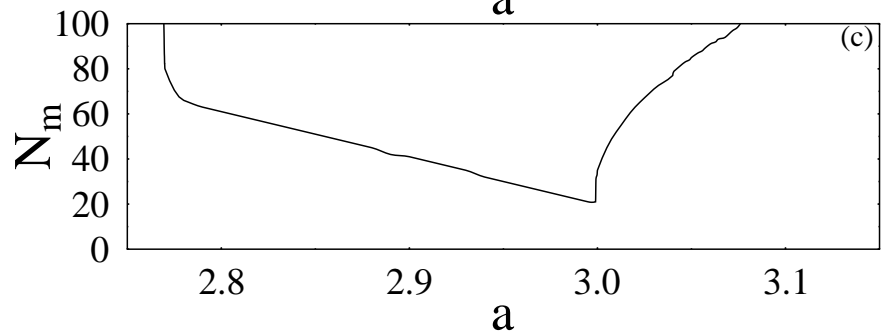
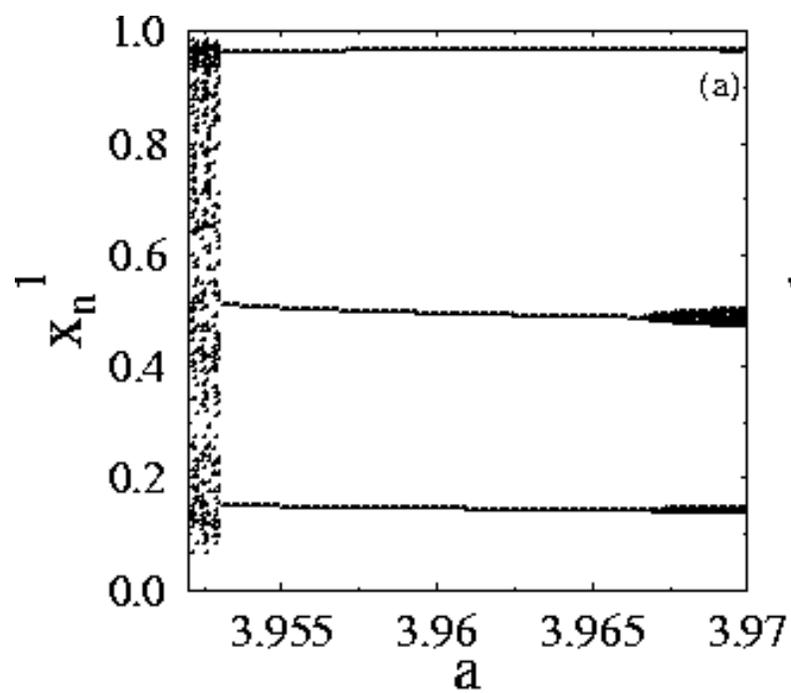
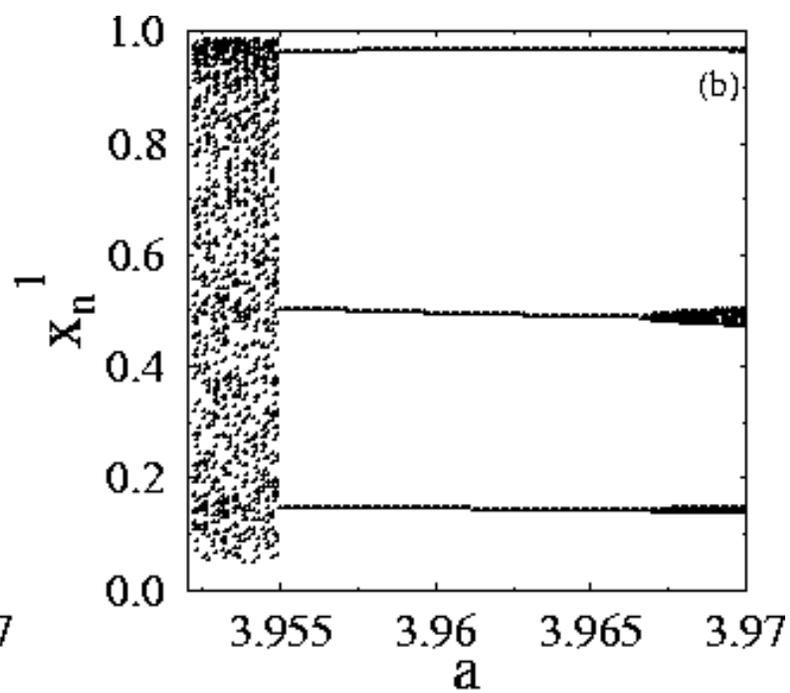
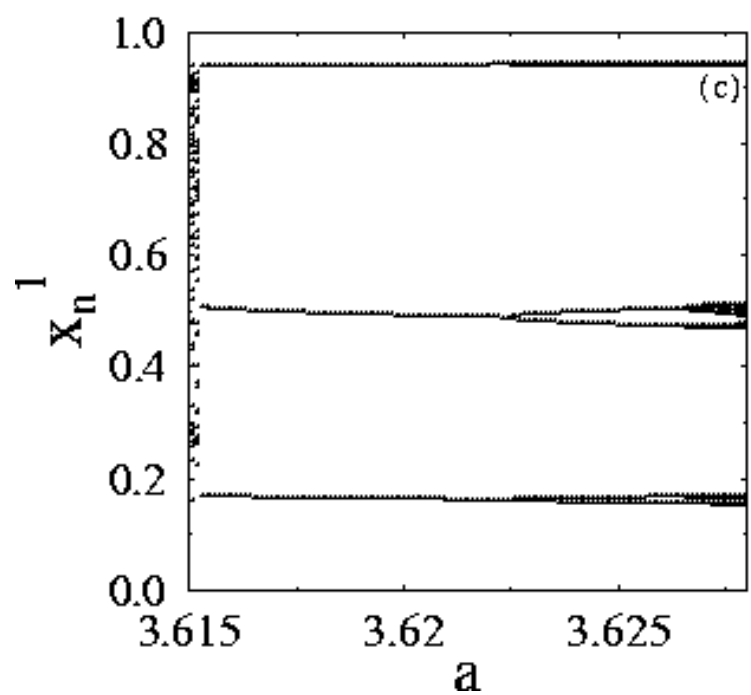
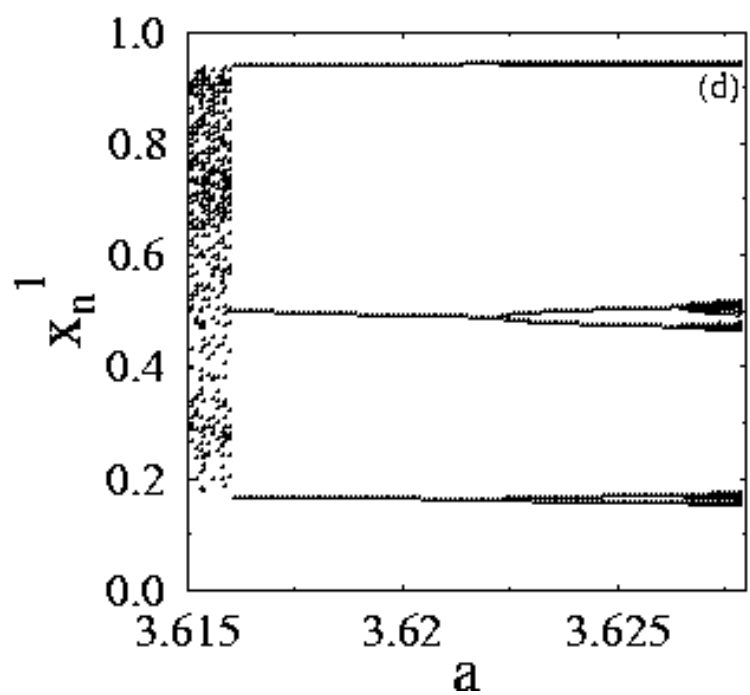
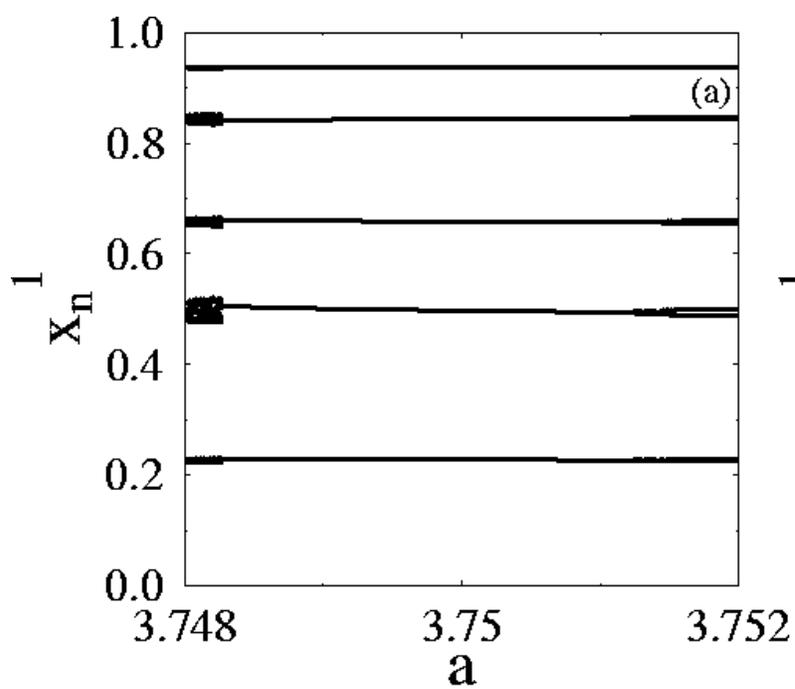
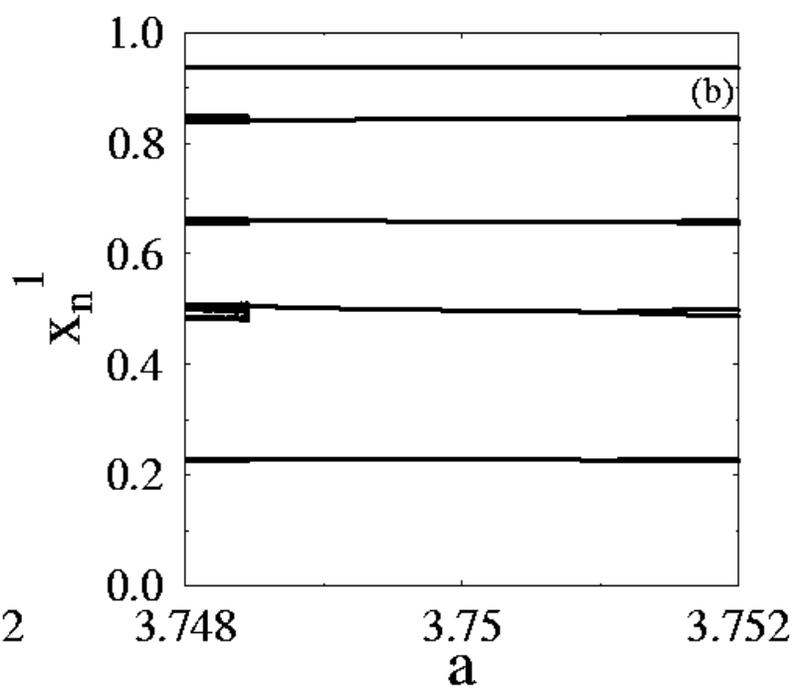
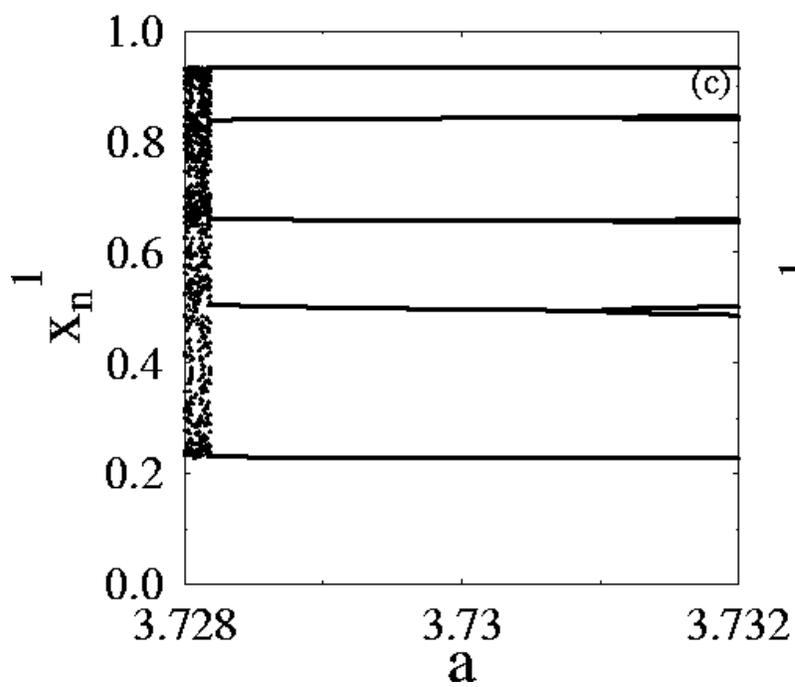
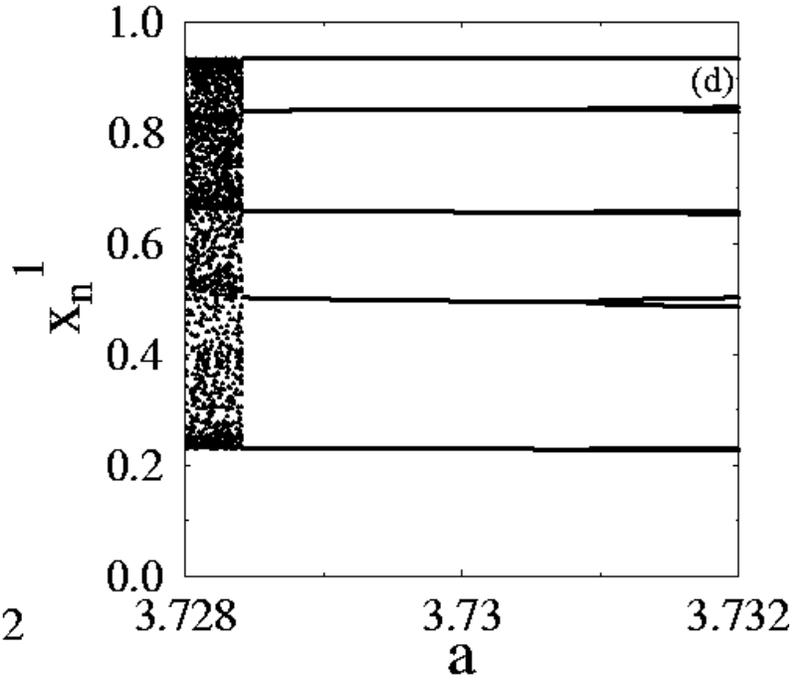
\end{document}